# Direct observation of charge separation in an organic light harvesting system by femtosecond time-resolved XPS

Friedrich Roth,[1,*] Mario Borgwardt[2], Lukas Wenthaus[3,4], Johannes Mahl[2], Steffen Palutke[4], Günter Brenner[4], Giuseppe Mercurio[5], Serguei Molodtsov[1,5,6], Wilfried Wurth[3,4,7,†], Oliver Gessner[2,*], and Wolfgang Eberhardt[3,*]

[1]Institute of Experimental Physics, TU Bergakademie Freiberg, D-09599 Freiberg, Germany

[2]Chemical Sciences Division, Lawrence Berkeley National Laboratory, Berkeley, California 94720, USA

[3]Center for Free-Electron Laser Science / DESY, D-22607 Hamburg, Germany

[4]Deutsches Elektronen-Synchrotron DESY, Notkestraße 85, 22603 Hamburg, Germany

[5]European XFEL GmbH, Holzkoppel 4, 22869, Schenefeld, Germany

[6]ITMO University, Kronverksky pr. 49, St. Petersburg, 197101, Russia

[7]Universität Hamburg, Luruper Chaussee 149, 22761, Hamburg, Germany

---

[*] corresponding authors, email: friedrich.roth@cfel.de, ogessner@lbl.gov, wolfgang.eberhardt@cfel.de
[†] Deceased

**Abstract**

The ultrafast dynamics of photon-to-charge conversion in an organic light harvesting system is studied by femtosecond time-resolved X-ray photoemission spectroscopy (TR-XPS) at the free-electron laser FLASH. This novel experimental technique provides site-specific information about charge separation and enables the monitoring of free charge carrier generation dynamics on their natural timescale, here applied to the model donor-acceptor system $CuPc:C_{60}$. A previously unobserved channel for exciton dissociation into mobile charge carriers is identified, providing the first direct, real-time characterization of the timescale and efficiency of charge generation from low-energy charge-transfer states in an organic heterojunction. The findings give strong support to the emerging realization that charge separation even from energetically disfavored excitonic states is contributing significantly, indicating new options for light harvesting in organic heterojunctions.

Photo-induced charge generation plays a central role in a broad range of physical, chemical, and biological processes that underlie natural and engineered photocatalytic and photovoltaic systems. Organic donor-acceptor systems are particularly intriguing candidates for light harvesting applications as their properties can be readily modified using well-established chemical synthesis techniques. Improving the efficiency of the underlying light-harvesting and charge generation processes, however, requires detailed knowledge of all the steps from the initial light-induced excitation of the chromophore to the final state where charges are separated in the donor and acceptor phases. $CuPc:C_{60}$ is a canonical model system for this class of devices but despite a significant body of research, fundamental mechanisms for charge separation remain obscure. Even more concerning, partly contradicting interpretations and models have been presented regarding the question which initial excitations contribute to charge generation and which do not. Evidently, a better theoretical understanding and novel experimental approaches are needed to validate or dismiss fundamental assumptions regarding the nature and fate of photo-excited states in organic heterojunctions.

Light harvesting in CuPc:$C_{60}$ is initiated through creation of an excitonic state at the chromophore (CuPc), while the desired final state consists of a separated electron-hole pair with a vacancy in the chromophore and a free electron in $C_{60}$. For many organic systems, $C_{60}$ is an excellent acceptor, capturing the electron and thus separating the charges. Adding a small amount of $C_{60}$, the fluorescence radiation from the recombination of the excitonic state of the chromophore is quenched[1,2], indicating a vastly improved efficiency of charge generation. Photoemission and inverse photoemission spectroscopy demonstrated that this process is energetically enabled by the electronic level alignment of the compounds forming the heterojunction, including CuPc and $C_{60}$ [3]. Let us review the knowledge about the energy landscape of the relevant electronic states involved. The energy level diagram of the CuPc:$C_{60}$ heterojunction, based on a combination of various spectroscopic data, is shown in Fig. 1. Singlet and triplet excitons in CuPc are located approximately 2 eV and 1.2 eV, respectively, above the ground state[3–7]. This is shown on the left of Fig. 1. Photoemission measurements established an offset of 1.45 eV between the HOMO of CuPc and $C_{60}$ [8]. This allows to determine the position of the HOMO of $C_{60}$ on the right side of Fig. 1 relative to the HOMO of CuPc. Combining photoemission and inverse photoemission spectroscopy, the onset of the HOMO-LUMO gap in solid $C_{60}$ was determined to 2.3 eV[9]. Gas-phase spectra of negatively charged $C_{60}^-$ indicate an energy difference of 2.0 eV between the $X^1A_g$ ground state and the $B^1S_1$ lowest singlet excited state of neutral $C_{60}$ [10]. Accordingly, the LUMO orbital of a fully occupied $C_{60}$ is located at an energy between 2.0 eV and 2.3 eV above the HOMO.

This corresponds to the charge-separated state configuration on the $C_{60}$ side of the heterojunction, where $C_{60}$ is now negatively charged. Finally, from the open circuit voltage of the CuPc:$C_{60}$ heterojunction solar cell, which corresponds to about 0.5 eV[11], we can derive a boundary for the minimum energy difference between the HOMO of CuPc and the LUMO of $C_{60}$.

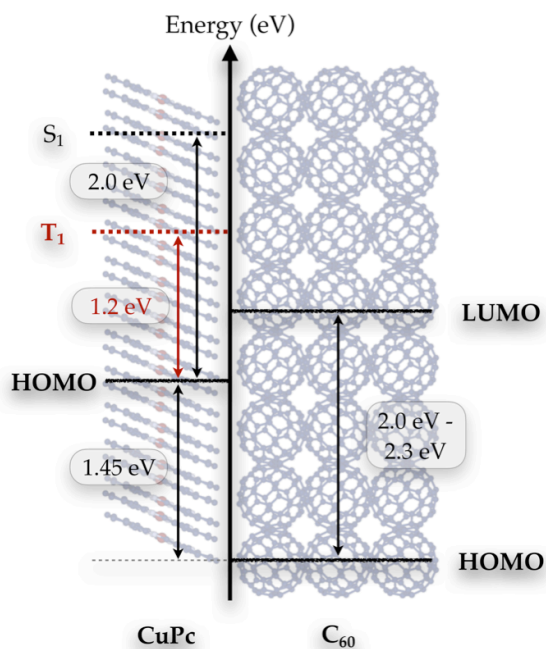

Fig. 1: Schematic energy level diagram of the CuPc:$C_{60}$ interface, based on values from the literature[3–6,13–16]. For more information see text.

While the (static) energetics is quite well established, the dynamics of the charge generation processes and the states involved are much less defined, if not to say controversial. As far as CuPc as chromophore is concerned, the lowest optically allowed excitation is a singlet exciton (S1) with approximately 2 eV excitation energy[3–7]. This exciton converts by intersystem crossing with a lifetime of about 500 fs into a triplet exciton (T1), which is located about 1.2 eV above the ground state [6,7] (cf. Fig. 1). When $C_{60}$ is in contact with CuPc, additional interface excitonic states (so-called interfacial charge transfer (ICT) states) with various electron and hole configurations exist [7], which are located several 100 meV in energy above the triplet state. These interface excitons are characterized by electronic configurations with the electron predominantly located in the $C_{60}$ while the hole remains in the adjacent CuPc. These ICT states can be viewed as precursors to the charge-separated state.

While a consensus has largely been reached regarding the static picture of the various states and their energy positions, their dynamics and the details of the processes leading to the generation of separate charges remain controversial. For example, a common notion found in the literature on CuPc:$C_{60}$ is that fully relaxed ICT [12] and triplet[5–7] states are too low in energy to result in separated charges, while opposition to this picture is more rare[9]. However, we previously showed[10] that T1 states indeed dominate the generation of free charges on timescales of 100's of picoseconds to nanoseconds. In contrast to these studies on Pc:$C_{60}$ systems, for a number of blends of $C_{60}$ with polymers or other small molecules, strong indications were found that low-energy charge-transfer excitons do in fact contribute to free carrier generation, even for the fully relaxed ICT excitons[13–15]. The direct, real-time observation of this charge generation channel, however, and its prevalence over competing loss channels is still outstanding.

Recently, we established Time-Resolved X-ray Photoelectron Spectroscopy (TR-XPS ) as a unique tool to detect the presence of charge transfer electrons in $C_{60}$ [16,17]. To the best of our knowledge, this is the first femtosecond time-resolved core-level photoemission spectroscopy study of Pc:$C_{60}$ systems. Access to the C 1s core-levels provides a unique perspective of the dynamic charge evolution in direct vicinity of the atom from which the core electron is emitted. This is the basis of Electron Spectroscopy for Chemical Analysis (ESCA), a widely-used standard tool for the study of the chemical and electronic structure of surfaces and interfaces. Ultrafast time-resolved XPS extends this capability into a new regime, allowing to monitor the femtosecond dynamics of charge generation and motion in complex interfacial systems with atomic specificity.

The technique is described in the methods section as well as in more detail in our previous publications[16,17]. Even though our earlier experiments were limited to a temporal resolution of ~70 ps, they clearly demonstrated that triplet states (T1) substantially contribute to the generation of separated charges in the CuPc:$C_{60}$ system. Integrated over all timescales from picoseconds to nanoseconds, the triplet states even generate an order of magnitude more charges than any other state[17]. Their long lifetimes more than compensate for their small diffusivities, leading to large diffusion lengths and efficient charge generation at the interface to the $C_{60}$.

Here, we extend this new spectroscopic technique to the femtosecond (fs) regime in order to capture the fast dynamics of the photoexcited states at the crucial moment when bound excitons dissociate into separated charges. We note that by carefully adjusting the fluence of the optical laser and the FEL, as well as the exposure time of each sample spot, we avoided complications such as sample charging or space-charge effects that affected previous PES investigations at FEL facilities. Specifically, we are interested in the relaxation of ICT states, which are considered important "gateways" for the generation of separate charges[12,18]. A common perception is that ICT states only dissociate into separate charges as hot states, but not when cooled down to the lowest energy of the corresponding state manifold[6,12]. Thus, the cooling time is considered a temporal gate for charge generation.

We use femtosecond TR-XPS in a pump-probe scheme at the free-electron laser FLASH to study a planar heterojunction consisting of a copper-phthalocyanine (CuPc) donor and a $C_{60}$ acceptor phase. FLASH offers femtosecond X-ray pulses with sufficiently high photon energies to study C 1s photoelectrons. As we have established previously[16,17], the binding energy of the C 1s core electrons of $C_{60}$ directly reflects the charge transfer on a local atomic scale. A shift of the C 1s line to lower binding energy indicates that additional electronic charge is present near the carbon atom that is ionized. Accordingly, XPS enables direct, local, and quantitative insight into the critically important step of the decay of the excitonic state into separated charges. Using an excitation wavelength of 775 nm, the experiment specifically addresses the relaxation dynamics of photo-induced ICT states near the low-energy limit of the ICT state manifold. TR-XPS spectra are recorded on timescales between ~100 fs and several picoseconds.

**Results**

Fig. 2 shows a series of time-resolved C 1s XPS spectra of a planar heterojunction consisting of 1 – 2 ML of CuPc atop a thin film of $C_{60}$ at four different pump-probe delays as indicated. Throughout this manuscript, the optical pump pulse precedes the X-ray probe pulse for positive pump-probe delays. The black solid line in Fig. 2a represents the ground state C 1s spectrum, since the optical pump pulse arrives 1 ps later than the X-Ray pulse (- 1ps). This spectrum can be described by a linear combination of the independently recorded C 1s spectra of CuPc (dashed) and $C_{60}$ (dotted)[3]. The peak located at a kinetic energy (KE) of 204 eV (within the yellow shaded area) is predominantly associated with $C_{60}$, whereas the structures outside the shaded area are photoemission signals from the carbon atoms in CuPc. The vertical gray line indicates the location of the $C_{60}$-C 1s line for the unperturbed, ground-state heterojunction. Fig. 2 b), c), and d) illustrate the photoinduced spectral dynamics at representative pump-probe delays of 0 ps, 0.5 ps, and 4 ps, respectively. Significant spectral changes are observed. In particular, the C 1s feature associated with $C_{60}$ shifts to higher kinetic energies and exhibits an overal line shape change.

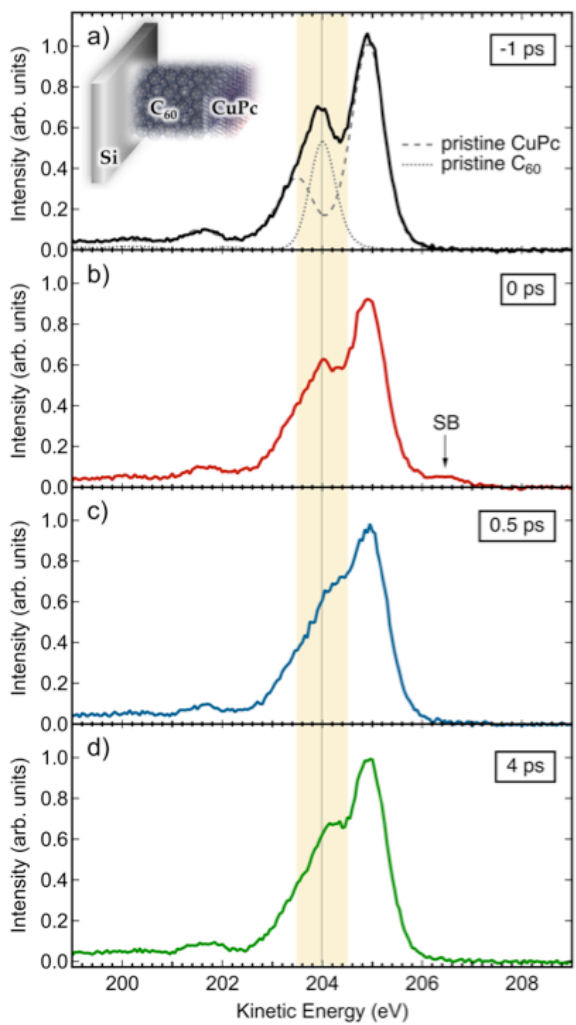

Fig. 2: Representative time-resolved C 1s-XPS spectra of a planar heterojunction consisting of ~1 – 2 ML of CuPc atop a thin film of $C_{60}$. Optical pump – X-ray probe time delays vary between -1 ps and 4 ps as indicated. In panel a), the spectra of pristine CuPc and $C_{60}$ are overlaid as dashed and dotted lines, respectively. The vertical gray line indicates the initial (ground state) position of the C 1s line of $C_{60}$, which dominates the spectrum within the yellow-shaded energy range. SB marks a sideband induced by the laser-assisted photoelectric effect near zero pump-probe delay.

Correspondingly, the minimum around 204.35 eV KE, between the $C_{60}$ peak and the main CuPc peak, becomes less pronounced and even vanishes with increasing pump-probe delay (Fig. 2 c).

These observations agree qualitatively with our previous, picosecond tr-XPS experiments on planar CuPc:$C_{60}$ heterojunctions[16,17] and extend them into the femtosecond regime. Additionally, a sideband feature (SB) appears at ≈ 206.5 eV KE when pump- and probe-pulses overlap in time (Fig. 2b) [19,20].

The shift of the $C_{60}$-C 1s photoline is interpreted as the signature of additional electronic charge in the vicinity of the atom being ionized and, accordingly, reflects directly upon the population of ICT states and free mobile electrons in the $C_{60}$ as result of donor-acceptor electron transfer. Interestingly, the low energy (201.65 eV) and high energy (204.95 eV) peaks associated with CuPc exhibit no significant shifts after optical laser excitation, in line with our previous results[16,17]. We note that the 775 nm pump laser photons are exclusively absorbed by the CuPc donors, while the $C_{60}$ acceptors do not exhibit any notable absorption at this wavelength . Moreover, neither a pure CuPc nor a pure $C_{60}$ film deposited on the same substrate as the heterojunction exhibits any photoinduced peak-shift aside from the rigid shift of the entire photoelectron spectrum due to the surface photovoltage effect in the Si support (see SM for details).

In order to analyze the dynamic trends quantitatively, a global fit is carried out based on a time-dependent decomposition of the heterojunction spectra into a linear combination of several pure component spectra associated with $C_{60}$ and CuPc. Fig. 3 shows a comparison between a) the measured data and b) the global fit for pump-probe delays between -2 ps and 4.4 ps using a 2D false-color representation as indicated by the color bar. Note that the sidebands near zero delay are included in the fit model.

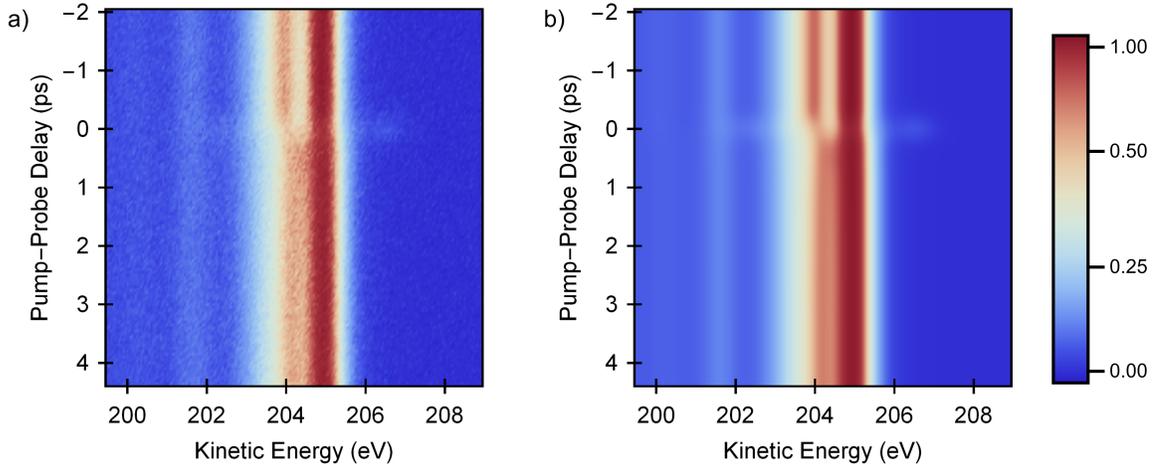

Fig. 3: (a) Measured femtosecond time-resolved C 1s-XPS spectra of the CuPc:$C_{60}$ heterojunction as a function of time-delay (vertical) and kinetic energy (horizontal). (b) Result of a global fit procedure that decomposes the heterojunction spectra into contributions associated with the two separate components CuPc and $C_{60}$.

As shown in more detail in Fig. 4 for some selected time delays, the TR-XPS spectra (black symbols) are modeled by a constant CuPc component (corresponding to the pristine CuPc spectrum in turquoise) and two $C_{60}$ components, $C_{60}(0)$ (purple) and $C_{60}(1)$ (red). The two $C_{60}$ components were set to have independent kinetic energies and amplitudes but the sum of their amplitudes is constant and corresponds to the amplitude of the $C_{60}$ component in the ground state spectrum. The two components represent $C_{60}$ molecules with ($C_{60}(1)$) and without ($C_{60}(0)$) additional screening charges in their vicinity. Correspondingly, the kinetic energy of $C_{60}(0)$ remains largely constant while $C_{60}(1)$ is shifted to higher kinetic energies by $\Delta E = 160 \pm 30$ meV relative to $C_{60}(0)$, which is interpreted as the result of an increased screening of the core hole by electrons in the vicinity of the core-ionized $C_{60}$ molecules. This is a dynamical screening process.

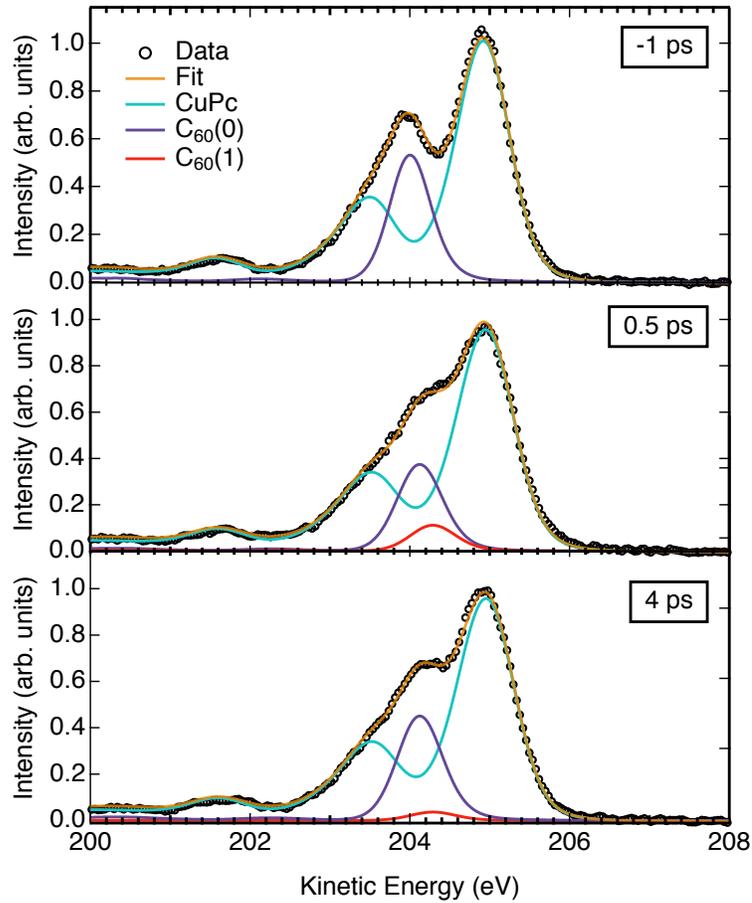

Fig. 4: Representative results of the global fit shown in Fig. 3 for three different pump-probe delays. Black symbols represent the measured C 1s spectra of the CuPc:$C_{60}$ heterojunction, orange lines the best fit. The turquoise, purple, and red solid lines show the fit components derived from the pristine material spectra as indicated. For positive delays, the total $C_{60}$ signal splits into an unperturbed component $C_{60}(0)$ and a new component $C_{60}(1)$ that is shifted to higher kinetic energies by ~160 meV. This shift of the $C_{60}(1)$-C 1s photoline is associated with increased screening of the core-hole by additional charge density in the vicinity of the core-ionized molecule.

If a C 1s atom is ionized, the local core hole is screened by the charges present in the vicinity. This screening charge is attracted not only within the $C_{60}$ molecule but also from neighboring $C_{60}$ molecules, since the electron hopping between $C_{60}$ molecules is a factor of 10 faster than the core hole lifetime. For comparison, we note that the $C_{60}$-C 1s binding energy is lowered by about 200 meV following intercalation of $C_{60}$ with alkali metal atoms, as in $K_1C_{60}$ or $Rb_1C_{60}$ [26,27]. The similarity of the shift values corroborates the physical interpretation of the $C_{60}(1)$ shift as the result of increased electronic screening of the core-ionized state by additional electron density compared to the $C_{60}(0)$ component, which results from core-ionization of $C_{60}$ without the presence of an additional charge. The time-dependence of the $C_{60}(1)$ signal, thus, provides direct, site-specific information about photoinduced charge population dynamics within the acceptor layer of the heterojunction. This specific information about the location of the photoinduced charge in the heterostructure is uniquely accessible through the TR-XPS measurement. For comparison, time-resolved transient absorption (TA)[28,29] and transient 2-photon valence band photoemission spectroscopy, which have been used previously to study these charge generation processes[5-8], allow to monitor the dynamics of various electronic states simultaneously. However, the experimental data only contain information on the energy scale of these states. Any information about the nature of these electronic states, their charge densities, and where these states are located in real-space has to be provided by theory. The relationship between theory and experimental data, however, is by no means trivial since standard theoretical techniques have known challenges to calculate the exact energies of excitonic states in these systems.

We note that, upon closer inspection of Fig. 4, a small transient shift of the $C_{60}(0)$ component to larger kinetic energies becomes apparent as well. This effect is ascribed to the finite precision of the fit procedure using the simplified two-component model to describe the transient spectral shifts of the $C_{60}$ C 1s spectrum. A more detailed description of the transient $C_{60}$-C 1s signals would need to take into account additional effects such as local environments within the $C_{60}$ film, which is beyond the scope of this work. Note, however, that in the following only the amplitudes and not the shifts of the $C_{60}(0)$ and $C_{60}(1)$ components are used to draw physical conclusions, which we consider robust against small variations in the absolute shift values. For completeness, we note that the absence of any measurable shifts in the CuPc C 1s peak may be related to both, the characteristic structure of the CuPc HOMO orbital and the corresponding HOMO hole, as well as different electron and hole mobilities in the donor and acceptor phases, leading to different final state contributions to the observed photoemission lines.

**Discussion**

Figure 5 a) shows the dynamic evolution of the $C_{60}(1)$ signal amplitude (red line). This XPS peak appears immediately, i. e., it rises within the instrument response function (IRF) of the experiment, and its amplitude decays in a bi-exponential fashion. This suggests that the population of electronically screened $C_{60}$ molecules appears immediately within the experimental response. Afterwards it initially decays very fast on a timescale of ~1 ps and then remains almost constant on timescales $\geq 25$ ps. The excitation wavelength (775 nm) is located in the low-energy fringe of the absorption spectrum of CuPc, which is associated with direct excitation of ICT states with the electron located in the $C_{60}$ and the hole in the CuPc[12]. According to Jailaubekov et al.[12], excitation at 800 nm results in the prompt appearance of an electric field at the interface due to direct ICT state excitation. Considering vibrational broadening and cooling within these states, there is no significant difference to be associated with our excitation energy, the center of which is 50 meV higher in energy; except that we have a slightly higher absorption in the sample. Consistent with this interpretation, the rise of the XPS signal associated with injected charges in the $C_{60}$ acceptor is instantaneous within our instrument response function. Thus, our study is an independent, strong indication for the direct excitation of ICT states also at 775 nm.

In contrast, excitation to higher-lying states leads to a delay of charge injection on the order of ~80-170 fs [6,12], which is not observed.

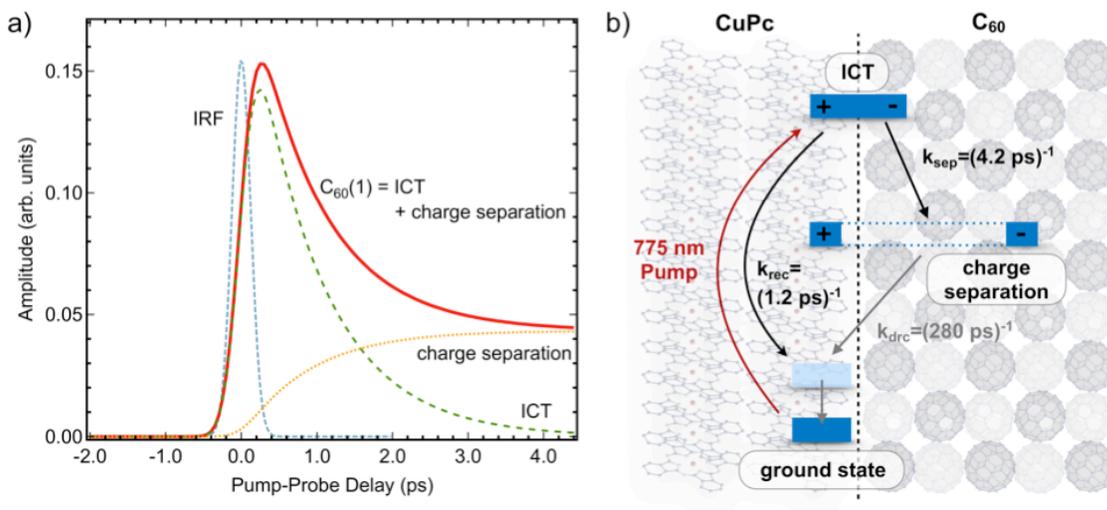

Fig. 5: a) Temporal evolution and b) kinetic modeling of the $C_{60}(1)$ peak intensity, representing the fraction of $C_{60}$ signal that is shifted toward smaller binding energies due to increased electronic screening. The blue curve in a) describes the instrument response function (IRF). The red $C_{60}(1)$ signal consists of a superposition of the fast decaying ICT signal (green) and a contribution from separate electron-hole pairs (yellow) that are generated during the decay of the ICT state.

Interestingly, the initial fast decay timescale of our $C_{60}(1)$ peak intensity is essentially identical to the decay timescale $\tau = 1.0 \pm 0.2$ ps of ICT states in planar CuPc-$C_{60}$ heterojunctions, reported by Jailaubekov et al.[12] and comparable to the timescale $\tau = 0.56 \pm 0.06$ ps determined by Dutton and Robey[6]. Importantly, the long-lived $C_{60}(1)$ contribution, represents a relaxation product of the initially prepared ICT states that, in contrast to the electron-hole recombination channel reported previously [6,12], results in a long-lived population of additional electrons within the $C_{60}$ acceptor phase. The instantaneous rise and smooth bi-exponential behavior of the intensity of the $C_{60}(1)$ peak is a strong indication that there are no other intermediate states involved. These observations are contrary to previous conclusions that the fully relaxed ICT states do not produce free carriers[6,12].

We speculate that a charge-separated signal component may have been present in previous experiments using valence-electron based spectroscopy techniques as well, but was not identified. This highlights the particular strengths of XPS as a quantitative technique that is ideally suited to study charge transfer processes. The atomic site-specificity paired with a deeper penetration depth as, for example, 2PPES, provides direct, unequivocal evidence for the long-lived charge-separated configurations that have evaded detection previously and are at the center of this study.

Additionally, core level XPS is a quantitative technique, since the cross-section for core-level photoemission does not depend on the chemical environment or the charge state. This is not the case for valence band 2PPES [5,8], where different states with different cross-sections are probed simultaneously, which impedes the interpretation of intensities in 2PPES spectra, especially when the probed states are overlapping in energy.

Note that for a related Pc system – ZnPc:$C_{60}$ blend films – Bartelt et al.[30] previously interpreted long-lived THz absorption signals as the result of free charge carrier generation in the $C_{60}$ domain following dissociation of ZnPc:$C_{60}$ ICT states after excitation by 800 nm radiation. In contrast to THz spectroscopy, TR-XPS is sensitive to the localization of electronic screening charge at core-ionized $C_{60}$ molecules from either bound ICT states or mobile charge carriers. The latter are of great importance for potential photovoltaic and photocatalytic applications as they provide a pathway to extract photogenerated charges from the heterojunction. More generally, the co-existence of two ICT relaxation channels, one "loss channel" (i. e., recombination) and one "use channel" (i. e., charge separation), is an ubiquitous concept to describe the efficiency of heterogeneous interface designs to convert photon energy into usable charge.

The TR-XPS results presented here provide a direct, quantitative probe of this efficiency as illustrated in Fig. 5. In order to obtain actual numerical data, the $C_{60}(1)$ signal in Fig. 4a) is decomposed into a time-dependent ICT (green, dashed) and a mobile electron (yellow, dotted) contribution using the global fit routine described above. The resulting model for photoexcitation and relaxation of the CuPc-$C_{60}$ heterojunction is displayed in Fig. 5b). Photoexcitation at 775 nm populates ICT states that decay either by charge recombination or by charge separation with rates of $k_{rec}$ = (1.2 ± 0.3 ps)$^{-1}$ and $k_{sep}$ = (4.2 ± 0.8 ps)$^{-1}$, respectively, corresponding to a net ICT relaxation rate of $k_{ICT}$ = (0.9 ± 0.2 ps)$^{-1}$. The ICT decay time is in excellent agreement with the results of Zhu and co-workers[18] and close to the value found by Dutton and Robey[6], providing additional support for the analysis presented here. The efficiency $\eta$ for mobile charge carrier generation from ICT states is directly determined by the ratio of the ICT relaxation channels: $\eta = k_{sep}/(k_{sep} + k_{rec}) = 0.22 \pm 0.07$. In other words, approximately 22 % of photogenerated ICT states relax into separate charge carriers that can be extracted from the heterojunction while the remaining ~78 % recombine and relax within ~1 ps. The charge-separated electron-hole pairs have a lifetime of 280 ps based on the results of our previous, picosecond time-resolved XPS study employing pump-probe delays up to nanoseconds[17].

The results presented here provide new insight into the photoexcited carrier dynamics on the femtosecond to picosecond timescale with direct impact on the photon-to-charge conversion efficiency of the CuPc-$C_{60}$ heterojunction. In contrast to previous conclusions[6,12], all interface excitonic states partially decay into separated charges, even for ICT states near the low-energy limit of the ICT band. This finding is consistent with the level scheme shown in Fig. 1 and calculations that locate the ICT states energetically above the triplet excitons in CuPc. As we have shown previously, the triplet states do contribute the vast majority of photogenerated charges for extended donor domains, albeit on rather slow, ~ ns timescales[17]. Thus, it is not surprising that the entire manifold of interface excitons, energetically located above the triplet states, contributes as well. Our results presented here are consistent with charge generation studies in a variety of other $C_{60}$ organic blends[13,14]. However, the femtosecond TR-XPS technique used here provides the first direct, real-time access to this important relaxation channel and its efficiency relative to competing loss mechanisms.

While a first principles treatment of the corresponding dynamics is still to be achieved, the results presented here provide critical benchmarks for renewed efforts toward a more complete theoretical description. In particular, the combination of detailed energetics, temporal dynamics, and relative channel efficiencies is ideally suited to support improved theoretical modelling. The direct determination of the ICT dissociation efficiency for an archetypical organic heterojunction described herein holds great promise that other light harvesting processes in complex, multi-component systems may be studied on their natural timescales and with the unprecedented site-specificity provided by ultrafast TR-XPS, paving a way toward a better understanding of emerging photovoltaic and photocatalytic frameworks.

**Methods**

*Time-resolved XPS measurements*

The TR-XPS experiment is carried out at the plane-grating monochromator beamline PG2 [31,32] at FLASH, using the photoemission end-station WESPE. A time-of-flight analyzer equipped with a segmented, position-sensitive delay-line detector is used to record the photoelectron spectra at a pass energy of 30 eV. FLASH is operated at a fundamental wavelength of 7.5 nm with an effective repetition rate of 4 kHz (pulse trains of 400 pulses each with an intra-train repetition rate of 1 MHz and a train repetition rate of 10 Hz) and a pulse duration of approximately 100 fs full-width-at-half-maximum (FWHM). A monochromator is used to select photon energies of ~ 495.8 eV from the third harmonic of the FLASH fundamental. The monochromator and electron analyzer settings are the same for all XPS measurements. Samples are excited by ~ 100 fs (FWHM) long optical laser pulses with a center wavelength of 775 nm, generated in an optical parametric chirped-pulse amplification (OPCPA) laser system that is operated with the same pulse pattern as FLASH[33]. Laser and X-ray focal spot sizes of 430 μm x 430 μm and 70 μm x 70 μm, respectively, ensure good spatial overlap and homogeneous excitation conditions across the probed sample area. The pump laser fluence on target is ~ 1 mJ/cm$^2$. The laser-assisted photoelectric effect (LAPE) of a Si (100) substrate [19,20] is used to adjust the temporal overlap between the optical pump- and X-ray probe-pulses and to determine the instrument response function (IRF), which is described by a Gaussian curve with a FWHM of 290 fs.

*Sample preparation*

The samples are prepared by in-situ sequential evaporation of $C_{60}$ and CuPc on the Si substrate from two spatially separated effusion cells. Precleaned and annealed n-type Si (100) wafers are used as substrates in order to ensure well-characterized, reproducible substrate conditions. Approximately two monolayers (MLs) of CuPc are deposited on top of a ~ 6 nm thick film of $C_{60}$. The film growth at room temperature is monitored by a quartz crystal microbalance to ensure homogeneous and continuous film deposition across the entire Si wafer. The samples are then transferred from the preparation chamber (base pressure < 3x10$^{-9}$ mbar) to the interaction chamber, which is operated at a base-pressure of 3 - 4x10$^{-10}$ mbar. Reference TR-XPS measurements are performed on the pure Si (100) substrate as well as, independently, on pristine, ~ 6 nm thick films of CuPc and $C_{60}$ deposited on the substrate (cf. SM for more details). In order to minimize the impact of sample damage by the X-ray and optical laser beams, a new sample spot is used every 10 minutes. The scan rate is calibrated by performing damage tests on sacrificial sample areas prior to recording the pump-probe data.

*Global Fit procedure and side band simulation*

A detailed description of the global fit procedure as well as the side band simulation can be found in the Supplemental Material.

**Data availability**

The data that support the findings of this study are available from the corresponding author upon reasonable request.

**Acknowledgments**

We would like to thank the staff at FLASH at DESY for their excellent support during the experiment. We especially thank H. Meyer and S. Gieschen for technical assistance. F.R. acknowledges financial support from DESY. This work was supported by the U.S. Department of Energy, Office of Science, Office of Basic Energy Sciences, Division of Chemical Sciences, Geosciences, and Biosciences at the Lawrence Berkeley National Laboratory under Contract No. DE-AC02-05CH11231. Moreover, this work was supported by the Helmholtz Associations Initiative and Networking Fund and Russian Science Foundation (Grant No. HRSF-0002/18-41-06001). M.B. acknowledges support by the Alexander von Humboldt foundation.

**Author contributions**

The experiment was conceived by W.E. and O.G., in discussion with F.R., L.W., W.W., and G.B. F.R., L.W., G.M. planned the experiment and prepared the samples. The experiments were performed by F.R., J.M., L.W., S.P., G.M. with input from W.E., O.G., W.W., and S.M. The data analysis and physical interpretation was performed by F.R., M.B., W.E. and O.G. All authors participated in discussing the data. The paper was written by F.R., O.G., and W.E. with input from all authors.

**Competing interests**

The authors declare no competing interest.

*Supplemental Material*

# Direct observation of charge separation in an organic light harvesting system by femtosecond time-resolved XPS

Friedrich Roth,[1*] Mario Borgwardt[2], Lukas Wenthaus[3], Johannes Mahl[2], Steffen Palutke[4], Günter Brenner[4], Giuseppe Mercurio[5], Serguei Molodtsov[1,5,6], Wilfried Wurth[3,4,7†], Oliver Gessner[2*], and Wolfgang Eberhardt[3*]

[1]Institute of Experimental Physics, TU Bergakademie Freiberg, D-09599 Freiberg, Germany

[2]Chemical Sciences Division, Lawrence Berkeley National Laboratory, Berkeley, California 94720, USA

[3]Center for Free-Electron Laser Science / DESY, D-22607 Hamburg, Germany

[4]Deutsches Elektronen-Synchrotron DESY, Notkestraße 85, 22603 Hamburg, Germany

[5]European XFEL GmbH, Holzkoppel 4, 22869, Schenefeld, Germany

[6]ITMO University, Kronverksky pr. 49, St. Petersburg, 197101, Russia

[7]Universität Hamburg, Luruper Chaussee 149, 22761, Hamburg, Germany

---

[*] corresponding authors, email: friedrich.roth@cfel.de, ogessner@lbl.gov, wolfgang.eberhardt@cfel.de
[†] Deceased

## I. TR-XPS measurements with pristine materials

Figures 1 and 2 show the time-dependent C 1s XPS signals from the two pristine materials, $C_{60}$ and CuPc, respectively, deposited on an n-type Si(100) substrate. No pump-induced changes are observed, except for the formation of sidebands due to the laser-assisted photoelectric effect (LAPE) during temporal overlap between the pump and probe pulses[1]. Furthermore, in both cases, a constant rigid shift of 30 meV of the entire photoelectron spectrum is observed, which is due to surface photovoltage effects in the Si substrate upon pumping with 775 nm light, and is included in the global fit approach.

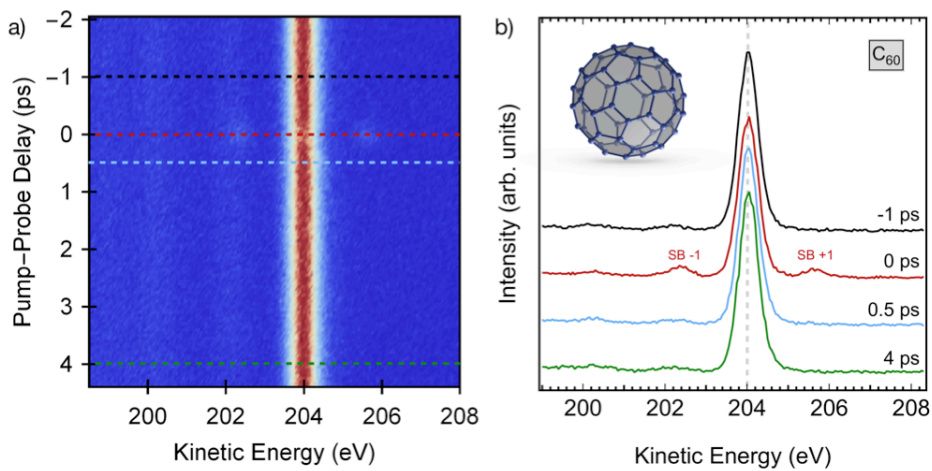

Fig. 1 a) 2D false-color map of the time-dependent C1s signal of a thin layer of $C_{60}$ (approx. 6 nm) on top of a pre-cleaned n-type Si wafer measured at a photon energy of 495.8 eV. Signal intensities increase from blue to red. b) Cuts at four different pump-probe delays as indicated in the false-color map.

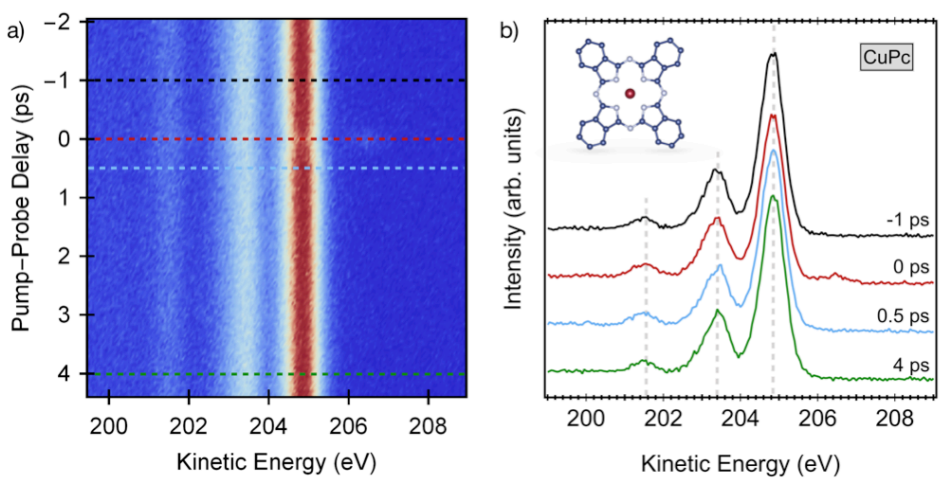

Fig. 2: a) 2D false-color map of the time-dependent C1s signal of a thin layer of CuPc (appr. 5.5 nm) on top of a pre-cleaned n-type Si wafer measured at a photon energy of 495.8 eV. Signal intensities increase from blue to red. Cuts at four different pump-probe delays as indicated in the false-color map.

## II. Detailed description of the global fit

Time-dependent changes in the XPS spectra of the CuPc-$C_{60}$ heterojunction are modeled based on a linear combination of the pure compound spectra ($C_{60}$ and CuPc). A global fit procedure is applied in order to reproduce the experimental data and to analyze the dynamic trends quantitatively in the time and energy domains. The linear combination is composed of three components: two pristine $C_{60}$ compound spectra, ($C_{60}(0)$ and $C_{60}(1)$) and a single pristine CuPc spectrum. For each component, the time-dependent kinetic energy (KE) and amplitude are free fit parameters, which account for all observed spectral shifts and intensity variations. The two $C_{60}$ components are necessary to describe the unperturbed and stationary contribution of $C_{60}$ (denoted by $C_{60}(0)$) and a second component (denoted as $C_{60}(1)$) accounting for the transient, additional charge present in the vicinity of some of the $C_{60}$ molecules. The time-dependence of the KEs are restricted by error functions for each basis component, reflecting the rigid spectral shifts induced by the surface photo-voltage effect. Time-zero, as well as the temporal width of the error functions are fixed to the result of the instrument response function (IRF) calibration measurements based on the LAPE effect as described in the main manuscript. This approach leads to two free fit parameters in the energy domain for each basis component, which are the characteristic position values before and after pump pulse interaction. To account for the splitting between the two $C_{60}$ contributions the energy offset between the two $C_{60}$ components is kept constant for all time delays, which reduces the number of free parameters by one. The dynamic evolution of the charge transfer process is modeled by a time dependent change in the amplitude ratio between the $C_{60}(0)$ and $C_{60}(1)$ under the restriction that their sum remains constant for all time delays. The amplitude of the $C_{60}(1)$ component ($A_{C_{60}(1)}$) is described by the sum of signals from the fast decaying ICT state $n_{ICT}$, populated by the 775 nm pump pulse, and contributions from charge-separated electronic states $n_{sep}$, which are populated via decay of the ICT state. Their transient populations are described as follows:

$$\frac{dn_{ICT}}{dt} = g(t - t_0, \sigma_{IRF}) - (k_{sep} + k_{rec})n_{ICT}(t) \qquad (1)$$

$$\frac{dn_{sep}}{dt} = k_{sep}n_{ICT}(t) - k_{src}n_{sep}(t) \qquad (2)$$

Here, $k_{sep}$ is the charge separation rate from the ICT state into a long-lived population, and $k_{rec}$ depicts the recombination rate from the ICT state back to the ground state (timescales and rates are related by $\tau_i = 1/k_i$). The decay of the charge-separated population occurs via the rate $k_{src} = (280 \text{ ps})^{-1}$, based on the results of a previous picosecond time-resolved XPS study [2] and fixed during the fit. $g(t)$ indicates a Gaussian-shaped initial population function accounting for the IRF. The two rates $k_{sep}$ and $k_{rec}$ are free fit parameters, while $g(t)$ is fixed to the measured IRF. We emphasize that the amplitude ratio between the two state populations is only determined by their decay rates and no additional weighting factors are applied. The amplitude of the $C_{60}(0)$ component is given by $A_{C60(0)} = A_{C60} - A_{C60(1)}$, i.e., the overall C60 signal amplitude $A_{C60}$ is kept constant for all time delays. The same restriction is applied to the CuPc component.

This global fit approach leads to a total of nine free fit parameters describing the entire 2D energy- and time-dependent transient signal. The global fit optimization was performed by minimizing the sum of the squares of the residuals when subtracting the linear combination of basis spectra from the measured transient signal. However, in the vicinity of time zero, the interaction of the X-ray emitted photoelectrons with the pump pulse leads to time-dependent intensity redistribution of the spectra into LAPE sidebands. To properly disentangle the system's internal photodynamics from this laser-induced effect, the simulation of the sideband formation is included in the fit procedure as detailed in the following section.

### III.  Side band simulations

The commonly used approach to quantitatively predict photoemission spectra in the presence of an intense laser field is a convolution of the unperturbed spectrum with a so-called LAPE response function. This function is constructed in order to simulate the partial redistribution of spectral intensities of the unperturbed spectrum into sidebands. The LAPE response for first order sidebands reads[1,3,4]:

$$f(E - E_0, \Delta t) = \frac{1 - 2A_1(\Delta t)}{\sqrt{2\pi\sigma^2}} e^{-(E-E_0)^2/2\sigma^2}$$
$$+ \sum_{\pm} \left( \frac{A_1(\Delta t)}{\sqrt{2\pi\sigma_1^2}} e^{-(E-E_0 \pm \hbar\omega)^2/2\sigma_1^2} \right)$$

Here, $A_1$ is the amplitude of the first sidebands, with a width $\sigma_1$ and a peak separation energy $\hbar\omega = 1.6\ eV$. The factor in front of the Gaussian peak is used to normalize the response function to 1. To simulate the dynamic response of the SB formation, the time-dependence of the amplitude ($A_1$) was fixed to a separately performed Gaussian fit of the sidebands region (IRF). At every iteration step of the global optimization procedure, the linear combination of the basis spectra was convoluted with the time-dependent LAPE response function.